\documentclass[aps,prd,twocolumn,showpacs,amsmath,amssym]{revtex4-1}
\usepackage{amsmath}
\usepackage{graphicx}
\usepackage{subfigure}
\usepackage{epstopdf}
\usepackage{color}
\usepackage{multirow}
\usepackage{setspace}
\usepackage{overpic}
\usepackage{amssymb}
\usepackage[dvipdfmx, bookmarksnumbered, pdfstartview=FitH,colorlinks,urlcolor=blue, citecolor=blue,linkcolor=blue] {hyperref}
\usepackage{lineno}
\usepackage{bm}
\usepackage{rotating}
\usepackage[utf8]{inputenc}
\let\oldequation\equation
\let\oldendequation\endequation

\renewenvironment{equation}
  {\linenomathNonumbers\oldequation}
  {\oldendequation\endlinenomath}

\begin{document}

\title{\bf \boldmath
Observation of $D^{0(+)}\to K^0_S\pi^{0(+)}\omega $ and improved measurement of $D^0\to K^-\pi^+\omega$
}

\author{
M.~Ablikim$^{1}$, M.~N.~Achasov$^{10,c}$, P.~Adlarson$^{67}$, S. ~Ahmed$^{15}$, M.~Albrecht$^{4}$, R.~Aliberti$^{28}$, A.~Amoroso$^{66A,66C}$, M.~R.~An$^{32}$, Q.~An$^{63,49}$, X.~H.~Bai$^{57}$
% ...
\\
\vspace{0.2cm}
(BESIII Collaboration)\\
\vspace{0.2cm} {\it
	$^{1}$ Institute of High Energy Physics, Beijing 100049, People's Republic of China\\
% ...
$^{4}$ Bochum Ruhr-University, D-44780 Bochum, Germany\\
% ...
$^{10}$ G.I. Budker Institute of Nuclear Physics SB RAS (BINP), Novosibirsk 630090, Russia\\
% ...
$^{15}$ Helmholtz Institute Mainz, Staudinger Weg 18, D-55099 Mainz, Germany\\
% ...
$^{28}$ Johannes Gutenberg University of Mainz, Johann-Joachim-Becher-Weg 45, D-55099 Mainz, Germany\\
% ...
$^{32}$ Liaoning Normal University, Dalian 116029, People's Republic of China\\
% ...
$^{49}$ State Key Laboratory of Particle Detection and Electronics, Beijing 100049, Hefei 230026, People's Republic of China\\
% ...
$^{57}$ University of Jinan, Jinan 250022, People's Republic of China\\
% ...
$^{63}$ University of Science and Technology of China, Hefei 230026, People's Republic of China\\
% ...
$^{66}$ University of Turin and INFN, (A)University of Turin, I-10125, Turin, Italy; (B)University of Eastern Piedmont, I-15121, Alessandria, Italy; (C)INFN, I-10125, Turin, Italy\\
	$^{67}$ Uppsala University, Box 516, SE-75120 Uppsala, Sweden\\
% ...
$^{c}$ Also at the Novosibirsk State University, Novosibirsk, 630090, Russia\\
% ...
}
}

\begin{abstract}
By analyzing an $e^+e^-$ annihilation data sample with an integrated luminosity of $2.93\ \rm fb^{-1}$ taken at the center-of-mass energy of 3.773~GeV with the BESIII detector, we determine the absolute branching fractions of the hadronic decays $D^0\to K^-\pi^+\omega$, $D^0\to K^0_S\pi^0\omega$, and
$D^+\to K^0_S\pi^+\omega$ to be
$(3.392 \pm 0.044_{\rm stat} \pm 0.085_{\rm syst})\%$,
$(0.848 \pm 0.046_{\rm stat} \pm 0.031_{\rm syst})\%$, and
$(0.707 \pm 0.041_{\rm stat} \pm 0.029_{\rm syst})\%$, respectively.
The accuracy of the branching fraction measurement of the decay $D^0\to K^-\pi^+\omega$ is improved by a factor of seven compared to the world average value.
The $D^{0}\to K^0_S\pi^{0}\omega$ and $D^{+}\to K^0_S\pi^{+}\omega$ decays are observed for the first time.
\end{abstract}


\maketitle

\oddsidemargin  -0.2cm
\evensidemargin -0.2cm

\section{Introduction}

Experimental studies of hadronic $D^{0(+)}$ decays can be used to investigate charm mixing, $CP$ violation, and strong-interaction effects~\cite{Asner:2008nq,Saur:2020rgd}.
Since the discoveries of the $D^{0(+)}$, hadronic $D^{0(+)}$ decays have been intensively studied~\cite{pdg2020}.
However, while the branching fraction of the decay $D^0\to K^-\pi^+\omega$ has previously been measured by the ARGUS Collaboration to be $(3.0\pm0.6)\%$~\cite{zpc56_7,pdg2020},
there are no measurements of the decays $D^0\to K^0_S\pi^0\omega$ and $D^+\to K^0_S\pi^+\omega$.
According to the statistical isospin model (SIM)~\cite{Rosner1,Rosner2}, the branching fractions for these isospin multiplets
are expected to satisfy
${ {\mathcal R}^0\equiv\frac{{\mathcal B}(D^0\to K^0_S\pi^0\omega)}{{\mathcal B}(D^0\to K^-\pi^+\omega)}=}0.4$ and
${ {\mathcal R}^+\equiv \frac{{\mathcal B}(D^+\to K^0_S\pi^+\omega)}{{\mathcal B}(D^0\to K^-\pi^+\omega)}=}0.9$.
Measurements of the branching fractions of these three decays can be used to test this relation.
In addition, these branching fractions provide important input for the background estimations needed to precisely test lepton flavor universality in semileptonic $B$ decays~\cite{lhcbnote} and to restrict new physics in the $D\to \bar K\pi\ell^+\ell^-$ decay~\cite{lhcb-kpimumu,babar-kpiee,bes3-kpiee}.

In this paper, we report an improved measurement of the branching fraction
of $D^0\to K^-\pi^+\omega$, and also report
the first observations and branching fraction measurements of the decays $D^0\to K^0_S\pi^0\omega$ and
$D^+\to K^0_S\pi^+\omega$.
 This analysis is performed by using an $e^+e^-$ collision data sample corresponding to an
integrated luminosity of 2.93~fb$^{-1}$~\cite{lum_bes3} collected at a
center-of-mass energy of $\sqrt s=$ 3.773~GeV  with the BESIII detector.
Throughout this paper, charge-conjugated modes are implied.

\section{BESIII detector and Monte Carlo simulation}
The BESIII detector is a magnetic
spectrometer~\cite{BESIII} located at the Beijing Electron
Positron Collider (BEPCII)~\cite{Yu:IPAC2016-TUYA01}. The
cylindrical core of the BESIII detector consists of a helium-based
 multilayer drift chamber (MDC), a plastic scintillator time-of-flight
system (TOF), and a CsI(Tl) electromagnetic calorimeter (EMC),
which are all enclosed in a superconducting solenoidal magnet
providing a 1.0~T magnetic field. The solenoid is supported by an octagonal flux-return yoke with
resistive plate modules interleaved with steel that provides muon identification. The acceptance of
charged particles and photons is 93\% of the solid angle. The
charged-particle momentum resolution at $1~{\rm GeV}/c$ is
$0.5\%$, and the $dE/dx$ resolution is $6\%$ for the electrons
from Bhabha scattering. The EMC measures photon energies with a
resolution of $2.5\%$ ($5\%$) at $1$~GeV in the barrel (end cap)
region. The time resolution of the TOF barrel region is 68~ps, while
that of the end cap region is 110~ps.
The effect of the trigger efficiency is negligible in this analysis according to Refs.~\cite{yy1,ZHU}.

Simulated data samples produced with a {\sc geant4}-based~\cite{geant4} Monte Carlo (MC) package, which includes the geometric description of the BESIII detector and the
detector response, are used to determine detection efficiencies
and estimate backgrounds. The simulation models the beam energy spread and initial state radiation (ISR) in the $e^+e^-$
annihilations with the generator {\sc kkmc}~\cite{kkmc}.
The inclusive MC samples include the production of $D\bar{D}$
pairs (including quantum coherence for the neutral $D$
channels), the non-$D\bar{D}$ decays of the $\psi(3770)$, the ISR
production of the $J/\psi$ and $\psi(3686)$ states, and the
continuum processes.
The known decay modes are modeled with {\sc
evtgen}~\cite{evtgen} using branching fractions taken from the
Particle Data Group (PDG)~\cite{pdg2020}, and the remaining unknown charmonium decays
are modeled with {\sc
lundcharm}~\cite{lundcharm}. Final state radiation
from charged final state particles is incorporated using the {\sc
photos} package~\cite{photos}.

\section{Double-tag method}

At $\sqrt s=3.773$ GeV,
the $\psi(3770)$ resonance is produced in $e^+e^-$ annihilation and decays predominantly into $D^0\bar D^0$ or $D^+D^-$ pairs,
without additional particles in the final state. This property allows us to determine absolute branching fractions
of $D$ decays with a double-tag method~\cite{Li:2021iwf}
.  In this method, the single-tag candidates are selected by reconstructing a $\bar D^0$ or $D^-$ in the following hadronic final states:
	$\bar D^0 \to K^+\pi^-$, $K^+\pi^-\pi^0$, and $K^+\pi^-\pi^-\pi^+$, and
	$D^- \to K^{+}\pi^{-}\pi^{-}$,
	$K^0_{S}\pi^{-}$, $K^{+}\pi^{-}\pi^{-}\pi^{0}$, $K^0_{S}\pi^{-}\pi^{0}$, $K^0_{S}\pi^{+}\pi^{-}\pi^{-}$,
	and $K^{+}K^{-}\pi^{-}$.
	The events in which a signal candidate can be reconstructed from the particles recoiling against the single-tag $\bar D$ meson are called double-tag events.
The branching fraction for the signal decay is determined by
\begin{equation}
\label{eq:br}
{\mathcal B}_{{\rm sig}} = N^{\rm tot}_{\rm DT}/(N^{\rm tot}_{\rm ST} \epsilon_{{\rm sig}}),
\end{equation}
where
$N^{\rm tot}_{\rm ST}=\sum_i N_{{\rm tag}}^i$ and $N^{\rm tot}_{\rm DT}$
represent the total single-tag yield and the double-tag yield, respectively; and
$\epsilon_{{\rm sig}}$ is the efficiency of detecting the signal decay in the presence of a single-tag $D$ meson,  averaged over all tag modes
\begin{equation}
\label{eq:eff}
\epsilon_{{\rm sig}} = \sum_i (N^i_{{\rm tag}} \epsilon^i_{{\rm sig}})/N^{\rm tot}_{\rm ST}.
\end{equation}
Here, $\epsilon^i_{{\rm sig}}=\epsilon^i_{{\rm tag,sig}}/\epsilon^i_{{\rm tag}}$,
where $\epsilon_{{\rm tag}}^i$ and $\epsilon_{{\rm tag,sig}}^i$
are the efficiencies of reconstructing the single-tag candidates and the double-tag events, respectively.

\section{Event selection}

We use the same selection criteria for $K^{\pm}$, $\pi^{\pm}$, $K_{S}^{0}$, $\gamma$ and $\pi^{0}$ as were used
in Refs.~\cite{epjc76,cpc40,bes3-pimuv,bes3-Dp-K1ev,bes3-etaetapi,bes3-omegamuv, bes3-kkpipi,bes3-etamuv,bes3-etax,bes3-dcs,bes3-benu}. The polar angle of each charged track is required to satisfy $|\rm{cos\theta}|<0.93$.
Except for those tracks used to reconstruct $K^0_{S}$ mesons,
their minimum distances to the interaction point~(IP) are required
to be within 10 cm in the beam direction and within 1 cm
in the perpendicular plane.
The particle identification~(PID) of each charged track is determined using
 $d E/d x$ and TOF information.
The combined confidence levels
for kaon and pion hypotheses are calculated, and denoted as
$CL_{K}$ and $CL_{\pi}$.
A charged track is identified as a kaon candidate if $CL_{K}$ is greater than
$CL_{\pi}$, or as a pion candidate if $CL_{\pi}$ is greater than
$CL_{K}$.

The $K^0_S$ candidates are obtained from two pions with opposite charges.
For the two charged pions, the requirement on
the closest approach to the IP in the beam direction is required to be within 20\,cm,
and no requirements on the distance in the perpendicular plane and on PID are imposed.
The two charged pions are constrained to originate from a common vertex at the IP, and
must have an invariant mass $M_{\pi^+\pi^-}$
within $\pm 12$~MeV/$c^2$ of the nominal mass of the $K^0_S$~\cite{pdg2020}.
For a $K^0_S$ candidate, the measured flight distance away from the IP is
required to be greater than twice its uncertainty.
Figure~1(a) shows the $M_{\pi^+\pi^-}$ distribution of $K^0_S$ candidates, where two solid arrows denote the $K^0_S$
signal region.

Photon candidates are chosen from isolated EMC clusters with energies greater than 25\,(50)\,MeV,
if the crystal with the maximum deposited energy in that cluster
is in the barrel~(end cap) region~\cite{BESIII}.
To reject photons from bremsstrahlung or from interactions with
material, showers within a $10^\circ$ cone around the
momentum direction of any charged track are rejected.
Reconstructed
showers due to electronic noise or beam backgrounds are
suppressed by requiring the timing information to be within $[0,\,700]$\,ns after the event start time.
The $\pi^0$ candidates are reconstructed from photon pairs.
To form a $\pi^0$ candidate,
the invariant mass of the photon pair is required to be in the interval $(0.115,\,0.150)$\,GeV$/c^{2}$.
A one-constraint~(1-C) kinematic fit is imposed on the photon pair to improve the momentum resolution.
In the kinematic fit, the invariant mass of the photon pair is constrained at
the nominal mass of the $\pi^{0}$~\cite{pdg2020}.

In the selection of $\bar D^0\to K^+\pi^-$ tags, the backgrounds
from cosmic rays and Bhabha events are suppressed by employing the
 following requirements.
First, the two charged tracks must have a
TOF time difference less than 5 ns, and they must not be
consistent with being a muon pair or an electron-positron
pair. Second, there must be at least one EMC shower
with an energy larger than 50 MeV or at least one additional
charged track detected in the MDC~\cite{deltakpi}.
In the selection of $\bar D^0\to K^+\pi^-\pi^-\pi^+$ candidates, the $\bar D^0\to K^0_SK^\pm\pi^\mp$ decays are suppressed by requiring the mass of all $\pi^+\pi^-$ pairs to
be outside of $(0.478,0.518)$~GeV/$c^2$.

\begin{figure*}[htbp]
  \centering
  \includegraphics[width=0.9\linewidth]{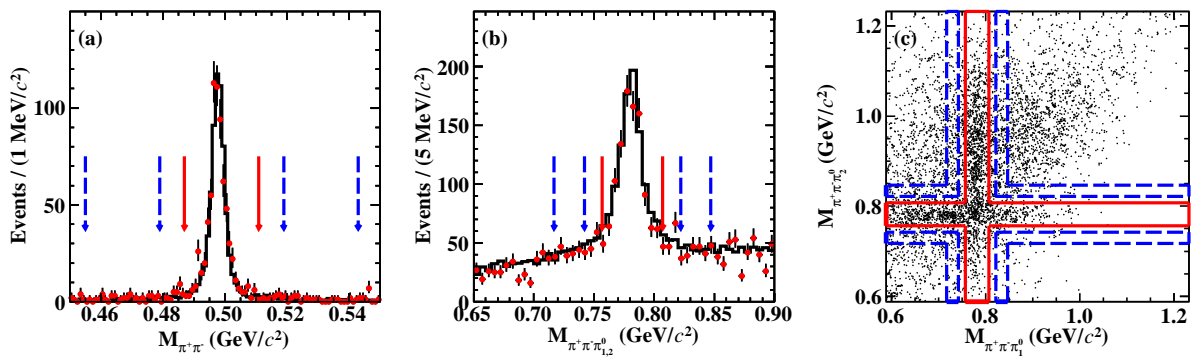}
  \caption{\small
(a)~The $M_{\pi^+\pi^-}$ distribution of the $K^0_S$ candidates,
(b) the combined $M_{\pi^+\pi^-\pi^0_1}$ and $M_{\pi^+\pi^-\pi^0_2}$ distribution of the $\omega$ candidates, and
(c) the distribution of $M_{\pi^+\pi^-\pi^0_1}$ versus $M_{\pi^+\pi^-\pi^0_2}$ from the $D^0\to K^0_S\pi^0\omega$ candidate events in data.
The red solid~(blue dashed) arrows [lines] show the corresponding boundaries of the {\color{blue}1D (2D) signal~(sideband)} region.
In these figures, except for the $K^0_S$ or $\omega$ mass requirement,
all other selection criteria and additional requirements of $|M^{\rm tag\,(sig)}_{\rm BC}-M_D|<0.005$~GeV/$c^2$ have been imposed.
The signal and sideband regions, illustrated here, are applied to all decays used in these analysis.
 In {\color{blue}plots} (a) and (b), points with error bars are data and histograms are the inclusive MC sample.
}\label{fig:sig}
\end{figure*}

To distinguish the tagged $\bar D$ (signal $D$) mesons from combinatorial backgrounds,
their energy difference and beam-constrained mass are defined as
\begin{equation}
\Delta E^{\rm tag\,(sig)} \equiv E_{\bar D\,(D)} - E_{\rm b},
\label{eq:deltaE}
\end{equation}
\begin{equation}
M_{\rm BC}^{\rm tag\,(sig)} \equiv \sqrt{E^{2}_{\rm b}/c^{4}-|\vec{p}_{\bar D\,(D)}|^{2}/c^{2}},
\label{eq:mBC}
\end{equation}
where tag and sig denote tag and signal sides, respectively;
$E_{\rm b}$ is the beam energy; and
$\vec{p}_{\bar D\,(D)}$ and $E_{\bar D\,(D)}$ are the momentum and energy of
the $\bar D\,(D)$ candidate in the rest frame of the $e^+e^-$ system.
For each tag (signal) mode, if multiple candidates are present,
the one with the smallest $|\Delta E^{\rm tag\,(sig)}|$ is kept for further analysis.
The $\Delta E^{\rm tag}$ is required to be in the range $(-55,\, +40)$\,MeV for the tag modes containing $\pi^0$
and within $(-25,\, +25)$\,MeV for the other modes.
The $\Delta E^{\rm sig}$ is required to be in the range $(-34,\, +41)$\,MeV, $(-42,\, +57)$\,MeV, and
$(-37,\, +44)$\,MeV for
$D^0\to K^-\pi^+\omega$, $D^0\to K^0_S\pi^0\omega$, and
$D^+\to K^0_S\pi^+\omega$, respectively.
These requirements correspond to about $3.5$ times the fitted mass resolutions away from the fitted peaks.
They are different for various modes to take different resolutions into account.
The requirements for the modes containing $\pi^0$ in the final state are asymmetric mainly
due to photon energy losses before entering the EMC.

The  $\omega$ candidates are selected from the $\pi^+\pi^-\pi^0$ combinations with an invariant mass
 $M_{\pi^+\pi^-\pi^0}$  within $\pm 25$ MeV/$c^2$  of the nominal $\omega$  mass~\cite{pdg2020}.
Figure~1(b) shows the $M_{\pi^+\pi^-\pi^0}$ distribution of $\omega$ candidates, where two solid arrows denote the $\omega$
signal region.
The
$D^{0}\to K^-\pi^{+}\omega$, $D^{0}\to K^0_S\pi^{0}\omega$ and $D^{+}\to K^0_S\pi^{+}\omega$ decays
are selected from the $K^-\pi^+\pi^+\pi^-\pi^0$, $K^0_S\pi^+\pi^+\pi^-\pi^0$ and $K^0_S\pi^+\pi^0\pi^-\pi^0$
final states.
As a result, there are two possible $\pi^+\pi^-\pi^0$  combinations
when selecting the $\omega$ candidate.
Their invariant masses are denoted as $M_{\pi^+_1\pi^-\pi^0}$ and $M_{\pi^+_2\pi^-\pi^0}$ for $K \pi^{+}\pi^+ \pi^- \pi^0$.
Their invariant masses are denoted as $M_{\pi^+\pi^-\pi^0_1}$ and $M_{\pi^+\pi^-\pi^0_2}$ for $K^{0}_{S} \pi^+ \pi^- \pi^0 \pi^0$, respectively.
Figure~\ref{fig:sig}(c)  shows the distribution of
$M_{\pi_2^+\pi^-\pi^0}$ versus $M_{\pi_1^+\pi^-\pi^0}$
for the accepted candidate events from the $D^0\to K^-\pi^+\omega$ decay in data.
Events in the two dimensional (2D) $\omega$ signal region,
 as outlined by the red lines in Fig.~\ref{fig:sig}(c),
are kept for further analysis.

\section{single-tag and double-tag yields}

\begin{figure}[htp]
  \centering
\includegraphics[width=1.\linewidth]{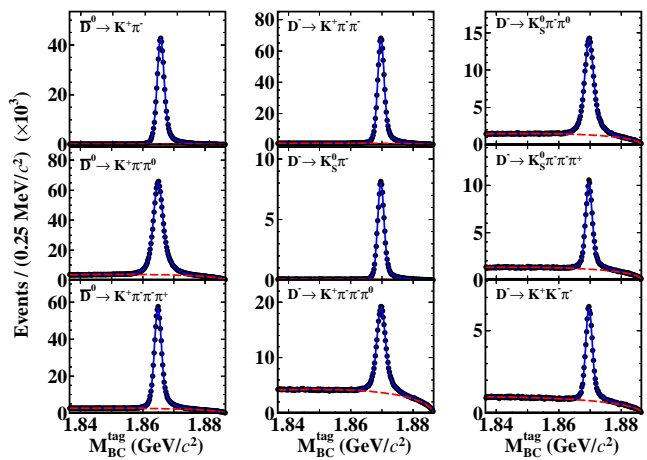}
  \caption{
Fits to the $M_{\rm BC}$ distributions of
the single-tag $\bar D^0$ (left column) and $D^-$ (middle and right columns) modes.
Data are shown as points with error bars.
The blue solid and red dashed curves are the fit results
and the fitted background shapes, respectively.
}
\label{fig:mBC_fit_data}
\end{figure}

The $M_{\rm BC}^{\rm tag}$ distributions of the accepted single-tag candidates in data
are shown in Fig.~\ref{fig:mBC_fit_data}. To determine the single-tag yield for each
tag mode, a maximum likelihood fit is performed on the corresponding $M_{\rm BC}^{\rm tag}$ spectrum.
An MC-simulated shape convolved with a double Gaussian function is used to model the $\bar D$ signal.
The convolved double Gaussian function describes the difference between the resolution in data and in MC simulation.
The combinatorial background shape is described by an ARGUS function~\cite{ARGUS}
defined as $c_f(f;E_{\rm end},\xi_f)=A_f\cdot f\cdot \sqrt{1 - \frac {f^2}{E^2_{\rm end}}} \cdot \exp\left[\xi_f \left(1-\frac {f^2}{E^2_{\rm end}}\right)\right]$,
where $f$ denotes $M^{\rm tag}_{\rm BC}$, $E_{\rm end}$ is an endpoint fixed at 1.8865 GeV, $A_f$ is a normalization factor, and $\xi_f$ is a free parameter.
Figure~\ref{fig:mBC_fit_data} shows the fit results on the $M_{\rm BC}^{\rm tag}$ distributions.
Finally, we obtain $1558163\pm2113$ single-tag $D^-$ mesons and $2386554\pm1928$ single-tag $\bar D^0$ mesons, denoted as $N^{\rm tot}_{\rm ST}$,
 where the uncertainties are statistical only.

\begin{figure}[htp]
\centering
\includegraphics[width=0.85\linewidth]{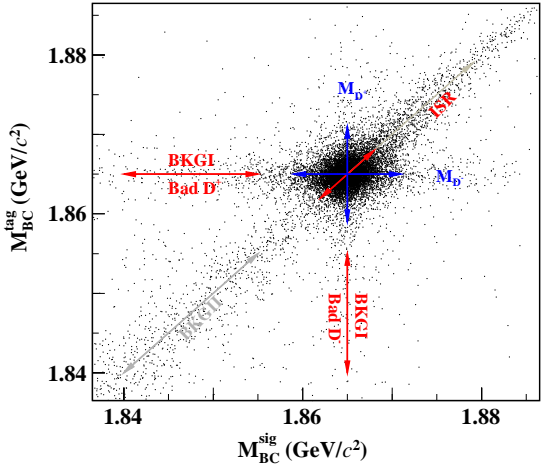}
\caption{
The distribution of $M_{\rm BC}^{\rm tag}$
versus $M_{\rm BC}^{\rm sig}$ of the accepted double-tag $D\bar D$ candidate events for $\bar D^0$ tags versus $D^0\to K^-\pi^+\omega$.
}
\label{fig:mBC2D}
\end{figure}

Figure~\ref{fig:mBC2D} illustrates the distribution of $M_{\rm BC}^{\rm tag}$
versus $M_{\rm BC}^{\rm sig}$ for double-tag candidate events.
Signal events concentrate around $M_{\rm BC}^{\rm tag} = M_{\rm BC}^{\rm sig} = M_{D}$,
where $M_{D}$ is the nominal $D$ mass~\cite{pdg2020}.
Background events are divided into three categories.
The first one (named BKGI) is from events with correctly reconstructed $D$ ($\bar D$) and incorrectly
reconstructed $\bar D$ ($D$), which are spread along the lines around
$M_{\rm BC}^{\rm tag}$ or $M_{\rm BC}^{\rm sig} = M_{D}$.
The second one (named BKGII) is from events smeared along the diagonal,
which are mainly from the $e^+e^- \to q\bar q$ processes.
The events with uncorrelated and incorrectly reconstructed $D$ and $\bar D$ (named BKGIII)
disperse across the whole allowed kinematic region. To extract the
double-tag yield in data, we perform a two-dimensional (2D) unbinned maximum likelihood
fit~\cite{cleo-2Dfit} on the underlying events. In the fit, the probability density functions (PDFs) of the
signal, BKGI, BKGII, and BKGIII are constructed as
\begin{itemize}
\item
signal: $a(x,y)$,
\item
BKGI: $b(x)\cdot c_y(y;E_{\rm b},\xi_{y}) + b(y)\cdot c_x(x;E_{\rm b},\xi_{x})$,
\item
BKGII: $c_z(z;\sqrt{2}E_{\rm b},\xi_{z}) \cdot g(k)$, and
\item
BKGIII: $c_x(x;E_{\rm b},\xi_{x}) \cdot c_y(y;E_{\rm b},\xi_{y})$,
\end{itemize}
respectively. Here, $x=M_{\rm BC}^{\rm sig}$, $y=M_{\rm BC}^{\rm tag}$, $z=(x+y)/\sqrt{2}$, and $k=(x-y)/\sqrt{2}$. The PDFs of signal $a(x,y)$, $b(x)$, and $b(y)$ are described by the corresponding MC-simulated shapes.
In addition, {\color{blue}$c_f(f;E_{\rm end},\xi_f)$} is an ARGUS function~\cite{ARGUS} (as defined above),
where $f$ denotes $x$, $y$, or $z$, and $E_{\rm b}$ is fixed at 1.8865 GeV; and
$g(k)$ is a Gaussian function with a mean of zero and a standard deviation parametrized by $\sigma_k=\sigma_0 \cdot(\sqrt{2}E_{\rm b}/c^2-z)^p$,
where $\sigma_0$ and $p$ are fit parameters.
For $x$ and $y$, $E_{\rm end}=E_b$, while $E_{\rm end}={\sqrt 2}E_b$ for $z$.

\begin{figure}[htp]
  \centering
\includegraphics[width=1.\linewidth]{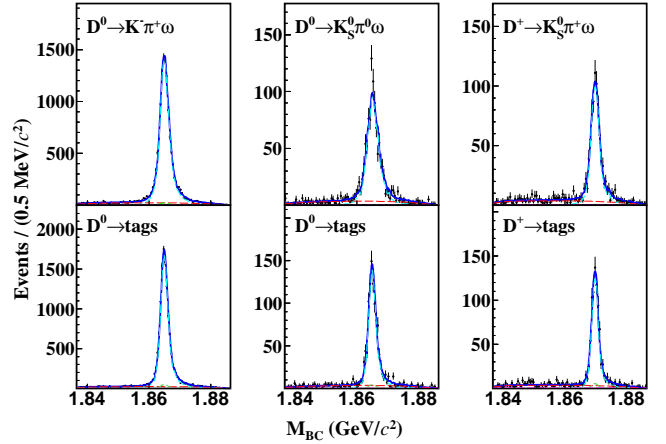}
\caption{Projections on $M^{\rm tag}_{\rm BC}$(bottom) and $M^{\rm sig}_{\rm BC}$(top) of the 2D fits to the double-tag candidate events for $D^0\to K^-\pi^+\omega$, $D^0\to K_{S}^{0}\pi^0\omega$, and $D^+\to K_{S}^{0}\pi^+\omega$.
	The dots with error bars are data. The blue curves are
	the fit results. The sky blue dashed curves are the fitted signal and the green curves and  light green dotted curves are
	the BKGI, the red curves are the BKGII, the purple curves are the BKGIII .
}
\label{fig:2Dfit}
\end{figure}

Projections on $M^{\rm tag}_{\rm BC}$ and $M^{\rm sig}_{\rm BC}$ of the 2D fits to the accepted double-tag candidate events
are shown in Fig.~\ref{fig:2Dfit}.
The second column of Table~\ref{table:br} presents
the double-tag yields for individual signal decays ($N^{\rm tot}_{\rm DT}$).

Potential peaking backgrounds are investigated as follows.
The combinatorial $\pi^+\pi^-$ pairs from the decays
$D^0\to K^-\pi^+\pi^+\pi^-\pi^0$, $D^0\to K^0_S\pi^+\pi^-\pi^0\pi^0$ and $D^+\to K^0_S\pi^+\pi^+\pi^-\pi^0$  or $\pi^+\pi^-\pi^0$ from the decays $D^0\to \pi^{+}\pi^-\pi^+\pi^-\pi^0\pi^0$ and $D^+\to \pi^+\pi^-\pi^+\pi^+\pi^-\pi^0$
 in the $K^0_S$ or $\omega$ signal region may survive in
the event selection criteria and form peaking backgrounds around
the $D$ mass in the $M^{\rm sig}_{\rm BC}$ distributions.
Figures~\ref{fig:sig}(a) and \ref{fig:sig}(c) also show the used 1D $K^0_S$ and 2D $\omega$ sideband regions, respectively.

In the studies of $D^{0(+)}\to K^0_S\pi^{0(+)}\omega$,
the candidate events lying in
the combined 1D $K^0_S$ sideband and 2D $\omega$ signal region,
the combined 1D $K^0_S$ signal and 2D $\omega$ sideband region, and
the combined 1D $K^0_S$ sideband and 2D $\omega$ sideband region,
are called peaking backgrounds A, B, and C, respectively.
In the study of $D^0\to K^-\pi^+\omega$, the candidate events in the 2D $\omega$ sideband region are called peaking background B.

The yields of the peaking backgrounds A/B/C ($N_{\rm sbA/B/C}$) are obtained by similar 2D fits to
the corresponding $M_{\rm BC}^{\rm tag}$ versus $M_{\rm BC}^{\rm sig}$ distributions of the individual candidate events  in data.
The net double-tag yield ($N^{\rm net}_{\rm DT}$) is calculated by
\begin{equation}
N^{\rm net}_{\rm DT}=N^{\rm tot}_{\rm DT}-f_1N_{\rm sbA}-f_2N_{\rm sbB}+f_1f_2N_{\rm sbC}.
\end{equation}
Here $f_{1}$ and $f_{2}$ are the factors normalizing the background yields in the $K^0_S$ and $\omega$ sideband regions to their
corresponding signal regions, respectively. These factors are obtained based on MC simulations.
The $f_1$ values are 0.5 for both $D^0\to K^0_S\pi^0\omega$ and $D^+\to K^0_S\pi^+\omega$ decays,
and it is taken to be 0 in the $D^0\to K^-\pi^+\omega$ decay because there is no $K^0_S$ involved.
The $f_2$ values are $1.44\pm0.04$, $1.32\pm0.03$, and $1.28\pm0.06$ for $D^0\to K^-\pi^+\omega$, $D^0\to K^0_S\pi^0\omega$ and $D^+\to K^0_S\pi^+\omega$, respectively,
where the uncertainties are mainly due to various sub-resonance components.
The values of $N_{\rm sbA}$, $N_{\rm sbB}$, $N_{\rm sbC}$, and $N^{\rm net}_{\rm DT}$ are summarized in Table~\ref{table:br}.

\section{Detection efficiencies and branching fractions}

\begin{figure*}[htbp]
  \centering
  \includegraphics[width=0.9\linewidth]{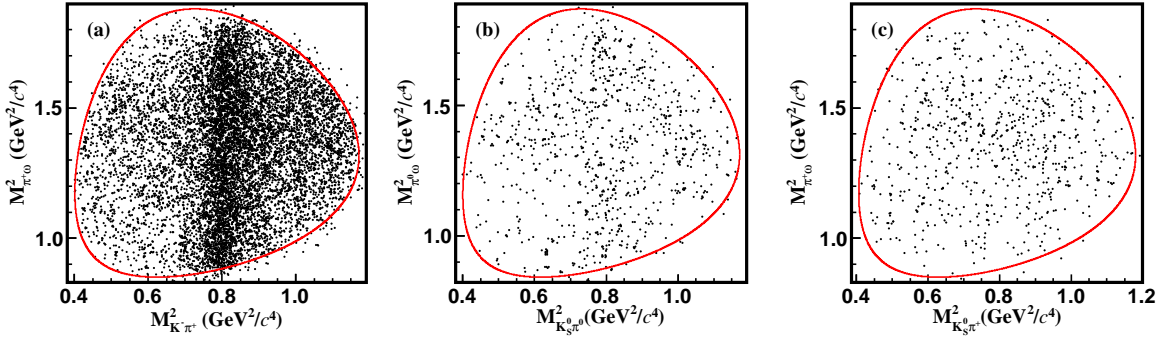}
  \caption{\small
  Dalitz plots of the accepted DT candidates for (a) $D^0\to K^-\pi^+\omega$,
  (b) $D^0\to K^0_S\pi^0\omega$, and (c) $D^+\to K^0_S\pi^+\omega$. The red curves show the kinematically allowed regions.
}\label{fig:Dalitz}
\end{figure*}

\begin{figure*}[htbp]
\centering
\includegraphics[width=0.85\linewidth]{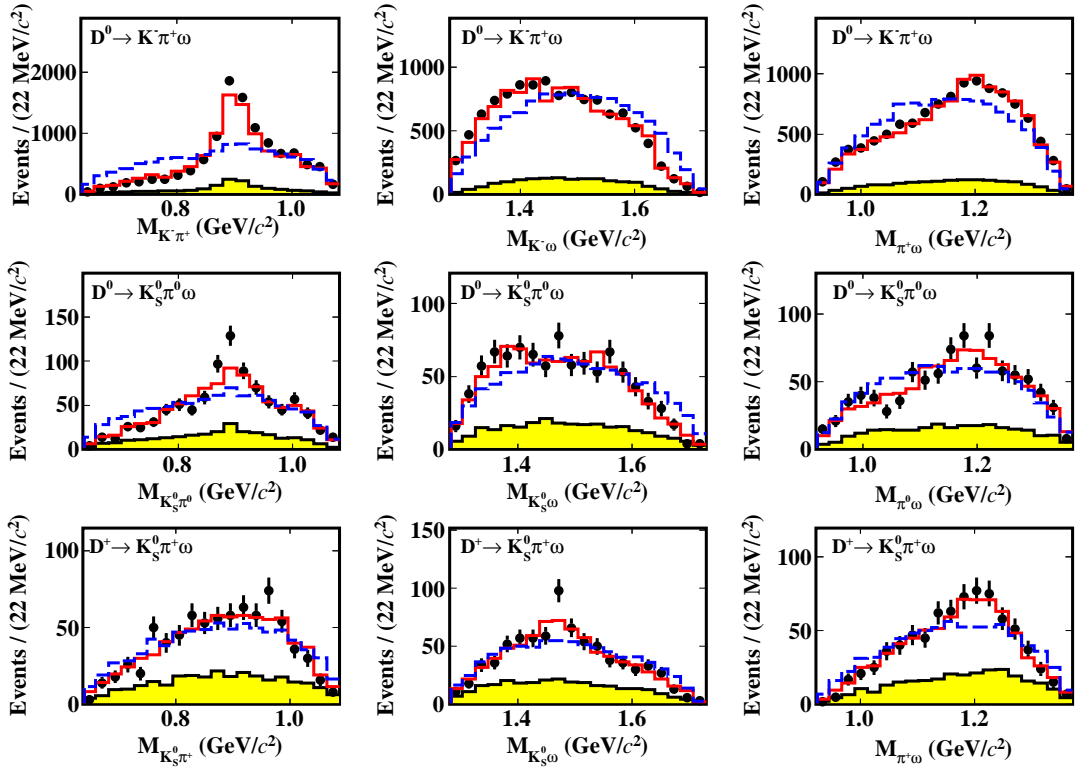}
\caption{
The $M_{\bar K\pi}$,
$M_{\pi\omega}$, and $M_{\bar K\omega }$ distributions of the $D\to\bar K\pi\omega$ candidate events
in the mass windows of $|M^{\rm tag\,(sig)}_{\rm BC}-M_D|<0.005$~GeV/$c^2$.
Data are shown as points with error bars;
the blue dashed and red solid histograms are the PHSP MC and modified MC samples, respectively;
the yellow shaded histograms are the simulated backgrounds estimated from the inclusive MC sample.}
\label{fig:mm_compare}
\end{figure*}

Figure~\ref{fig:Dalitz} shows the Dalitz plots of the accepted candidates in data.
Figure~\ref{fig:mm_compare} displays the $M_{\bar K\pi}$, $M_{\pi\omega}$, and $M_{\bar K\omega}$ distributions
of $D\to \bar K \pi\omega$ candidate events after requiring $|M_{\rm BC}^{\rm tag\,(sig)}-M_D|<0.005$ GeV/$c^2$.
 Signals of $\bar K^{*0}(892)$ can be observed in the $M_{\bar K\pi}$ spectra for $D^0\to K^-(K^0_S)\pi^{+(0)}\omega$ decays.
However, no obvious $K^{*+}(892)$ signal is observed in the $M_{\bar K\pi}$ spectrum for the $D^+\to K^0\pi^+\omega$ decay
because the $D^+\to K^{*+}(892)\omega$ decay is a doubly Cabibbo-suppressed process.
We can see that the distributions of MC simulated events,  uniformly generated across the phase space,
are not in good agreement with the data distributions.
To resolve these disagreements, we modify the MC generator
according to the Dalitz plots of $D\to \bar K\pi\omega$ in data, by implementing the same method as used in Ref.~\cite{bes3_kpieta}.
In the Dalitz plot, the signal and background components are modeled by the efficiency-corrected phase space (PHSP) MC simulation and
the inclusive MC simulation, respectively.
In Fig.~\ref{fig:mm_compare}, we can see that the modified MC distributions agree well with the data distributions.
Therefore, we use the modified MC samples to evaluate the detection efficiencies
of the signal decays ($\epsilon_{\rm sig}$). The resulting efficiencies are summarized in the seventh column of Table~\ref{table:br}.

With the numbers of $N^{\rm tot}_{\rm ST}$, $N^{\rm net}_{\rm DT}$, and $\epsilon^{}_{{\rm sig}}$ shown in Table~\ref{table:br},
the branching fractions of the hadronic decays $D^0\to K^-\pi^+\omega$, $D^0\to K^0_S\pi^0\omega$, and $D^+\to K^0_S\pi^+\omega$ are determined. The resulting branching fractions are summarized in the eighth column of Table~\ref{table:br}.

\begin{table*}[htp]
\centering
\caption{\label{table:br}
\small
The double-tag yields in data ($N^{\rm tot}_{\rm DT}$),
the yields of peaking backgrounds A/B/C ($N_{\rm sbA/B/C}$),
the net double-tag yields in data ($N^{\rm net}_{\rm DT}$),
the signal efficiencies ($\epsilon_{\rm sig}$), the branching fractions ($\mathcal B_{\rm sig}$) and
the comparisons to world averages ($\mathcal B_{\rm PDG}$).
The efficiencies include the branching fractions of $K^0_S\to \pi^+\pi^-$,
$\omega\to\pi^+\pi^-\pi^0$, and $\pi^0\to \gamma\gamma$.
The uncertainties of $N^{\rm tot}_{\rm DT}$
and $\epsilon_{\rm sig}$ are statistical only, the first and second uncertainties of
$\mathcal B_{\rm sig}$ are statistical and systematic, respectively.
}
\begin{tabular}{lcccccccc}
\hline\hline
Decay mode             & $N^{\rm tot}_{\rm DT}$ & $N_{\rm sbA}$& $N_{\rm sbB}$& $N_{\rm sbC}$& $N^{\rm net}_{\rm DT}$& $\epsilon_{\rm sig}$\,(\%)& $\mathcal B_{\rm sig}$\,(\%)& $\mathcal B_{\rm PDG}$\,(\%)\\ \hline
$D^0\to K^-\pi^+\omega$  &$11745\pm115$&...&$1092\pm38$&...&$10174\pm128$&$12.96\pm0.06$&$3.392 \pm 0.044 \pm 0.085 $&$3.0\pm0.6$ \\
$D^0\to K^0_S\pi^0\omega$&$944\pm33$&$27\pm6$&$70\pm10$&$190\pm14$&$697\pm41$&$3.43\pm0.04$&$0.848 \pm 0.046 \pm 0.031$&N/A \\
$D^+\to K^0_S\pi^+\omega$&$770\pm30$&$28\pm6$&$67\pm10$&$130\pm13$&$523\pm35$&$5.35\pm0.04$&$0.707 \pm 0.041 \pm 0.029$&N/A \\
\hline\hline
\end{tabular}
\end{table*}

\section{Systematic uncertainties}
\label{sec:sys}

In the measurements of the branching fractions (${\mathcal B}(D^0\to K^{-}\pi^{+}\omega)$, ${\mathcal B}(D^0\to K^{0}_{S}\pi^{0}\omega)$, and ${\mathcal B}(D^+\to K^{0}_{S}\pi^{+}\omega)$) and the branching fraction ratios (${\mathcal R}^0\equiv\frac{{\mathcal B}(D^0\to K^0_S\pi^0\omega)}{{\mathcal B}(D^0\to K^-\pi^+\omega)}$ and ${\mathcal R}^+\equiv \frac{{\mathcal B}(D^+\to K^0_S\pi^+\omega)}{{\mathcal B}(D^0\to K^-\pi^+\omega)}$),
the sources of the systematic uncertainties are summarized in Table~\ref{table:sys}.
Each of them is evaluated relative to the measured branching fraction.
See below for details.

\begin{table*}[htp]
\centering
\caption{\label{table:sys} Relative systematic uncertainties (in \%) in the determinations of ${\mathcal B}(D^0\to K^{-}\pi^{+}\omega)$, ${\mathcal B}(D^0\to K^{0}_{S}\pi^{0}\omega)$, ${\mathcal B}(D^+\to K^{0}_{S}\pi^{+}\omega)$, ${\mathcal R}^0$, and ${\mathcal R}^+$.
For the last two columns, the numbers before and after `/' are the uncanceled systematic uncertainties of
${\mathcal B}(D^0\to K^{-}\pi^{+}\omega)$ and ${\mathcal B}(D^{0(+)}\to K^{0}_{S}\pi^{0(+)}\omega)$
in calculating ${\mathcal R}^{0}$ and ${\mathcal R}^{+}$.     }
\begin{small}
\begin{tabular}{lccccc}
\hline
Source & ${\mathcal B}(D^0\to K^{-}\pi^{+}\omega)$&${\mathcal B}(D^0\to K^{0}_{S}\pi^{0}\omega)$&${\mathcal B}(D^+\to K^{0}_{S}\pi^{+}\omega)$&${\mathcal R}^0$ & ${\mathcal R}^+$\\
\hline
  Single-tag yields              & 0.5  & 0.5 & 0.5 & -- &0.7  \\
  Tracking of $K^\pm$            & 0.3  & --  & --  &0.3 &0.3  \\
  Tracking of $\pi^\pm$          & 0.9  & 0.6 & 0.9 &0.3 &-- \\
  PID of $K^\pm$                 & 0.3  & --  & --  &0.3 &0.3  \\
  PID of $\pi^\pm$               & 0.9  & 0.6 & 0.9 &0.3 &-- \\
  $K_S^0$ reconstruction         &  --  & 1.5 & 1.5 &1.5&1.5  \\
  $\pi^0$ reconstruction         & 0.7  & 1.4 & 0.7 &0.7 &-- \\
  2D $M_{\rm BC}$ fit            & 0.5  & 1.5 & 1.8 &1.6&1.9 \\
  $\Delta E$ requirement         & 0.3  & 0.4 & 0.4 &0.5&0.5 \\
  MC modeling                    & 0.6  & 1.2 & 2.1 &2.2&1.3 \\
  MC statistics                  & 0.4  & 0.7 & 0.6 &0.8&0.7 \\
  Quoted branching fractions     & 0.8  & 0.8 & 0.8 &0.1&0.1 \\
  Quantum correlation effect     & 0.6   & 0.4 & --   &0.7&0.6  \\
  $K^0_S$ sideband               & --   & 0.1 & 0.4 &0.1&0.4  \\
  $\omega$ sideband              & 1.2  & 1.0 & 1.3 &1.6&1.8 \\
  Rejection of $D\to\bar K\pi\eta$ & 0.5  & 1.2 & 0.3 &1.3&0.6 \\
  The factor of $f^\omega$       & 0.3  & 0.7 & 1.0 &--&-- \\
  \hline
  Total & {\color{blue}2.5} & 3.7 & 4.1 & 4.0 & 3.8 \\
  \hline
\end{tabular}
\end{small}
\end{table*}

\begin{itemize}
\item
{\bf Single-tag yields:}
The uncertainties in the total single-tag $\bar D^0$ and $D^-$ yields were
studied in~\cite{epjc76,cpc40,bes3-pimuv}.
They are 0.5\% for both neutral and charged $\bar D$.

\item
{\bf  Tracking (PID) of $K^\pm/\pi^\pm$}:
The tracking (PID) efficiencies of $K^\pm/\pi^\pm$
are investigated with double-tag $D\bar D$ hadronic events using a
partial reconstruction (identification) technique. No significant biases are found.
The averaged ratios between data and MC efficiencies ($f_{K\,{\rm or}\,\pi}^{\rm tracking\,(PID)}=\epsilon_{K\,{\rm or}\,\pi}^{\rm tracking\,(PID)}[{\rm data}]/\epsilon_{K\,{\rm or}\,\pi}^{\rm tracking\,(PID)}[{\rm MC}]$) of tracking (PID) for $K^\pm$ or $\pi^\pm$ are weighted by the corresponding momentum spectra of signal MC events,
giving $f_K^{\rm tracking}$ to be 1.021 and $f_\pi^{\rm tracking}$ to be close to unity. After correcting the MC efficiencies by $f_K^{\rm tracking}$,
the residual uncertainties of $f_{K\,{\rm or}\,\pi}^{\rm tracking}$ are assigned as the systematic uncertainties of tracking efficiencies,
which are 0.3\% per $K^\pm$ and 0.3\% per $\pi^\pm$. $f_K^{\rm PID}$ and $f_\pi^{\rm PID}$ are all close to unity and their individual uncertainties, 0.3\% are taken as the associated systematic uncertainties per $K^\pm$ or $\pi^\pm$.

\item
{\bf\boldmath $K_S^0$ reconstruction}:
A control sample of $J/\psi\to K^{*}(892)^{\mp}K^{\pm}$ and $J/\psi\to \phi K_S^{0}K^{\pm}\pi^{\mp}$
was used to evaluate the efficiency of the $K_{S}^{0}$ reconstruction~\cite{sysks}.
The efficiency includes effects due to the $\pi^+\pi^-$ track selection, the vertex fit, requirements on the
decay length, and the $K^0_S$ mass window. We take 1.5\% per $K^0_S$ to be
the associated systematic uncertainty.

\item
{\bf\boldmath $\pi^0$ reconstruction}:
To study the efficiency of $\pi^0$ reconstruction, we used the control sample of the double-tag
$D\bar D$ hadronic decays $D^0\to K^-\pi^+$, $K^-\pi^+\pi^+\pi^-$ versus
$\bar D^0\to K^+\pi^-\pi^0$, $K^0_S\pi^0$~\cite{epjc76,cpc40}.
The efficiency includes effects due to the photon selection, the 1-C kinematic fit, and the $\pi^0$ mass window.
The data/MC ratio for the $\pi^0$ reconstruction efficiency is found to be $(99.1\pm0.7)\%$. After correcting the MC signal efficiency by this ratio, a  systematic uncertainty of 0.7\% is assigned to each $\pi^0$.

\item
{\bf\boldmath 2D fit on $M^{\rm tag}_{\rm BC}$ versus $M^{\rm sig}_{\rm BC}$}:
To investigate the systematic uncertainty due to the 2D fit on $M^{\rm tag}_{\rm BC}$ versus
$M^{\rm sig}_{\rm BC}$, we remeasure the branching fraction by using alternative signal shapes with different MC
matching requirements, and alternative endpoints of the ARGUS function $E_{\rm b}\pm0.2$\,MeV/$c^2$.
For each source, the largest changes to the nominal branching fractions are taken as its systematic uncertainty.
Quadratically summing over these two effects, we assign the systematic uncertainties in the 2D fit
to be 0.3\%, 1.2\%, and 2.1\% for $D^0\to K^-\pi^+\omega$, $D^0\to K^0_S\pi^0\omega$, and
$D^+\to K^0_S\pi^+\omega$, respectively.

\item
{\bf\boldmath  $\Delta E_{\rm sig}$ requirement}:
The systematic uncertainty due to the $\Delta E_{\rm sig}$ requirement is assigned to be 0.3\%, 0.4\%, and 0.4\% for
$D^0\to K^-\pi^+\omega$, $D^0\to K^0_S\pi^0\omega$, and
$D^+\to K^0_S\pi^+\omega$, respectively,
which corresponds to the largest efficiency difference with and without smearing the data-MC Gaussian resolution of $\Delta E_{\rm sig}$ for signal MC events.
Here, the smeared Gaussian parameters are obtained by using the samples of double-tag events $D^0\to K^-\pi^+$,  $D^0\to K^-\pi^+\pi^0$, $D^0\to K^-\pi^+\pi^0\pi^0$, $D^+\to K^0_S\pi^+$ and $D^+\to K^-\pi^+\pi^+\pi^0$ versus the same $\bar D$ tags in our nominal analysis.

\item
{\bf\boldmath MC modeling}:
To estimate the systematic uncertainty in the MC modeling, we vary
the bin sizes and the MC-simulated background sizes of the input
$M^2_{\bar K\pi}$ versus $M^2_{\pi\omega}$ distributions in the generator by one quarter.
The largest changes to the nominal detection efficiencies,
0.6\%, 1.2\%, and 2.1\%, are taken as the homologous systematic uncertainties for
$D^0\to K^-\pi^+\omega$, $D^0\to K^0_S\pi^0\omega$, and
$D^+\to K^0_S\pi^+\omega$, respectively.

\item
{\bf\boldmath MC statistics}:
The  uncertainties due to MC statistics are considered as a source of
systematic uncertainties, which are 0.4\%, 0.7\%, and 0.6\% for
$D^0\to K^-\pi^+\omega$, $D^0\to K^0_S\pi^0\omega$, and
$D^+\to K^0_S\pi^+\omega$, respectively.

\item
{\bf\boldmath Quoted branching fractions}:
The uncertainties of the quoted branching fractions of $K^0_S\to \pi^+\pi^-$,
$\omega\to\pi^+\pi^-\pi^0$, and $\pi^0\to \gamma\gamma$
 are also considered as a source of systematic uncertainties, which are
0.07, 0.8\%, and 0.03\%~\cite{pdg2020}, respectively.

\item
{\bf\boldmath Quantum correlation effect}:
The measurement of the branching fraction of the neutral $D$ decay is affected by the quantum correlation effect. For the decay $D^0\to K^0_S\pi^0\omega$  the $CP$-even  component is estimated by the $CP$-even tag $D^{0}\to K^{+}K^{-}$ and the $CP$-odd tag $D^0\to K^{0}_{S}\pi^{0}$. Using the same method as described in Ref.~\cite{QC-factor} and the necessary parameters quoted from Refs.~\cite{R-ref1,R-ref2,R-ref3}, the correction factor to account for the quantum correlation effect on the measured branching fraction is found to be  (96.1$\pm$0.4$_{\rm stat}$)\%. After correcting the signal efficiency by this factor, the residual error, 0.4\%, is  assigned as the associated systematic uncertainty.
The associated uncertainty in the measurement of the branching fraction of $D^0 \to K^-\pi^+\omega$
is assigned to be 0.6\% by the method discussed in Ref.~\cite{cleo-qc}, and
it is controlled by the ratio of Cabibbo-suppressed and Cabibbo-favored rate combined
with the strong phase difference between two amplitudes.

\item
{\bf\boldmath $K^0_S$ sideband, $\omega$ sideband, and rejection of $D\to \bar K\pi\eta$}:
The uncertainties due to the choices of the $K^0_S$ and $\omega$ sideband regions (the rejection requirement of $D\to \bar K\pi\eta$) are studied by shifting the nominal regions (requirement) by $\pm 0.005\,(0.010)$ GeV/$c^2$. For each source, the largest difference of the branching fraction is taken as the corresponding systematic uncertainty for each signal decay.

\item
{\bf\boldmath The factor of $f^{\omega}$}:
The systematic uncertainty due to the factor $f^\omega$ is assigned by examining the changes of branching fractions via varying the factors by $\pm1\sigma$. They are assigned as 0.3\%, 0.7\% and 1.0\% for $D^0\to K^-\pi^+\omega$, $D^0\to K^0_S\pi^0\omega$ and $D^+\to K^0_S\pi^+\omega$, respectively.

\end{itemize}

Adding these systematic uncertainties in quadrature, we obtain the total systematic uncertainties for the measurements of
${\mathcal B}(D^0\to K^-\pi^+\omega)$, ${\mathcal B}(D^0\to K^0_S\pi^0\omega)$, ${\mathcal B}(D^+\to K^0_S\pi^+\omega)$,
${\mathcal R^0}$, and ${\mathcal R^+}$ to be 2.5\%, 3.7\%, 4.1\%, 4.0\%, and 3.8\%, respectively.

\section{Summary}

With a data sample corresponding to an integrated luminosity of $2.93\,\rm fb^{-1}$,
taken with the BESIII~\cite{white} detector at $\sqrt{s}=3.773$\,GeV, we measure
the absolute branching fractions of $D^0\to K^-\pi^+\omega$, $D^0\to K^0_S\pi^0\omega$, and
$D^+\to K^0_S\pi^+\omega$.
The branching fractions of $D^0\to K^0_S\pi^0\omega$ and
$D^+\to K^0_S\pi^+\omega$ are measured for the first time.
Our result of the branching fraction of $D^0\to K^-\pi^+\omega$ is consistent
with the previous result within uncertainties,
but the accuracy is improved by a factor of seven.
The comparisons of measurements and the world-average values  are shown in Table~\ref{table:br}. Using the branching fractions measured in this work, we determine the branching fraction ratios to be
${ {\mathcal R}^0\equiv\frac{{\mathcal B}(D^0\to K^0_S\pi^0\omega)}{{\mathcal B}(D^0\to K^-\pi^+\omega)}=}0.23\pm0.01_{\rm stat}\pm0.01_{\rm syst}$ and
${ {\mathcal R}^+\equiv \frac{{\mathcal B}(D^+\to K^0_S\pi^+\omega)}{{\mathcal B}(D^0\to K^-\pi^+\omega)}=}0.21\pm0.01_{\rm stat}\pm0.01_{\rm syst}$.
Both ${\mathcal R^0}$ and ${\mathcal R^+}$ significantly deviate from their expected values of the SIM.
These deviations may arise from a possible strong phase difference between two decay amplitudes due to final state interactions~\cite{Rosner1,Rosner2}.

\section{Acknowledgements}
The BESIII collaboration thanks the staff of BEPCII and the IHEP computing center for their strong support. This work is supported in part by National Key Research and Development Program of China under Contracts Nos. 2020YFA0406400, 2020YFA0406300; National Natural Science Foundation of China (NSFC) under Contracts Nos. 11625523, 11635010, 11775230, 11735014, 11822506, 11835012, 11935015, 11935016, 11935018, 11961141012; the Chinese Academy of Sciences (CAS) Large-Scale Scientific Facility Program; Joint Large-Scale Scientific Facility Funds of the NSFC and CAS under Contracts Nos. U1732263, U1832207, U1632109; CAS Key Research Program of Frontier Sciences under Contracts Nos. QYZDJ-SSW-SLH003, QYZDJ-SSW-SLH040; 100 Talents Program of CAS; INPAC and Shanghai Key Laboratory for Particle Physics and Cosmology; ERC under Contract No. 758462; European Union Horizon 2020 research and innovation programme under Contract No. Marie Sklodowska-Curie grant agreement No 894790; German Research Foundation DFG under Contracts Nos. 443159800, Collaborative Research Center CRC 1044, FOR 2359, FOR 2359, GRK 214; Istituto Nazionale di Fisica Nucleare, Italy; Ministry of Development of Turkey under Contract No. DPT2006K-120470; National Science and Technology fund; Olle Engkvist Foundation under Contract No. 200-0605; STFC (United Kingdom); The Knut and Alice Wallenberg Foundation (Sweden) under Contract No. 2016.0157; The Royal Society, UK under Contracts Nos. DH140054, DH160214; The Swedish Research Council; U. S. Department of Energy under Contracts Nos. DE-FG02-05ER41374, DE-SC-0012069.

\end{document}